\documentclass[lettersize,journal]{IEEEtran}
    \usepackage{booktabs}
    \usepackage{diagbox}
    \usepackage{makecell}
    \usepackage{array}
    \usepackage{graphicx,amssymb,amsmath}
    \usepackage{multicol}
    \usepackage[noadjust]{cite}
    \usepackage{setspace}
    \usepackage{subfigure}
    \usepackage{graphicx}
    \usepackage{float}
    \usepackage {url}
    \usepackage{stfloats}
    \usepackage{amsthm,pifont}
    \usepackage{flushend}
    \usepackage{cases,subeqnarray}
    \usepackage{bm,multirow,bigstrut}
    \usepackage{amsmath, amsthm, amssymb}
    \usepackage{textcomp}
    \usepackage{latexsym,bm}
    \usepackage{booktabs}
    \usepackage{xcolor}
    \usepackage{mathtools}
    \usepackage{dsfont}
    \usepackage{extarrows}
    \usepackage{epsfig}
    \usepackage{epsfig}
    \usepackage{epstopdf}
    \usepackage{hyperref}

    \theoremstyle{plain}

    \theoremstyle{plain}

    \usepackage{amsmath}

    \IEEEoverridecommandlockouts

\begin{document}
\title{Optimizing 6G Integrated Sensing and Communications (ISAC) via Expert Networks}
\author{Jiacheng Wang, Hongyang Du, Geng Sun, Jiawen Kang, Haibo Zhou, Dusit~Niyato,~\IEEEmembership{Fellow,~IEEE}, and Jiming Chen,~\IEEEmembership{Fellow,~IEEE}
\thanks{J.~Wang, H.~Du and D. Niyato are with the School of Computer Science and Engineering, Nanyang Technological University, Singapore (e-mail: jiacheng.wang@ntu.edu.sg, hongyang001@e.ntu.edu.sg, dniyato@ntu.edu.sg).}

\thanks{G. Sun is with the College of Computer Science and Technology, Jilin University, Changchun, China (e-mail: sungeng@jlu.edu.cn).}

 \thanks{J. Kang is with the School of Automation, Guangdong University of Technology, China (e-mail: kavinkang@gdut.edu.cn).}

  \thanks{H. Zhou is with the School of Electronic Science and Engineering, Nanjing University, Nanjing, China (haibozhou@nju.edu.cn).}

    \thanks{J. Chen is with the College of Control Science and Engineering, Zhejiang University, Hangzhou, China (jmchen@ieee.org).}

}

\maketitle

\begin{abstract}
Integrated Sensing and Communications (ISAC) is one of the core technologies of 6G, which combines sensing and communication functions into a single system. However, limited computing and storage resources make it impractical to combine multiple sensing models into a single device, constraining the system's function and performance. Therefore, this article proposes enhancing ISAC with the mixture of experts (MoE) architecture. Rigorously, we first investigate ISAC and MoE, including their concepts, advantages, and applications. Then, we explore how MoE can enhance ISAC from the perspectives of signal processing and network optimization. Building on this, we propose an MoE based ISAC framework, which uses a gating network to selectively activate multiple experts in handling sensing tasks under given communication conditions, thereby improving the overall performance. The case study demonstrates that the proposed framework can effectively increase the accuracy and robustness in detecting targets by using wireless communication signal, providing strong support for the practical deployment and applications of the ISAC system.
\end{abstract}
\IEEEpeerreviewmaketitle
\section{Introduction}
Integrated Sensing and Communication (ISAC) is a novel paradigm in the realm of wireless networks. It leverages advancements in solid-state circuits, microwave, and wireless technologies to combine communication and sensing operations on a single platform, through shared antenna arrays, signal processing algorithms, and networking protocols. Such an integration not only enhances the efficiency in resource utilization, such as spectrum, hardware, and energy, but also supports the development of emerging technologies, such as autonomous driving~\cite{liu2022integrated}. Therefore, ISAC stands to revolutionize the way wireless systems are designed and deployed, making it a more adaptive and resource-efficient wireless ecosystem.

An ISAC system is capable of sensing various aspects of targets while simultaneously supporting data communication. For instance, the authors in~\cite{wang2024generative} employed different algorithms, such as fast Fourier transform (FFT) and generative diffusion model, to process down-link signals and extract features in the time, frequency, and spatial domains. Then, by analyzing the obtained features through clustering, fine-grained human flow detection was achieved based on wireless communication signals. For a particular task, the ISAC system requires the use of various models. These models include traditional algorithms for signal processing, discriminative AI (DAI) for data analysis and classification, and generative AI (GAI) for data enhancement and generation. However, in practice, users' demands are unpredictable, and deploying and executing multiple models on a single ISAC device is not feasible due to limited storage and computing capabilities. As a result, ISAC encounters challenges in practical applications.

\begin{itemize}
    \item Given that a limited type and number of models can be deployed on a single ISAC system, for the tasks requested by users, the system may not have corresponding models to handle them directly.

    \item ISAC systems are equipped with models capable of processing requested tasks, but the communication rate may not meet the model's requirements, which makes them unable to operate effectively.
\end{itemize}

For instance, in WiFi sensing, a user requests dynamic target detection at a data packet transmission rate of 200 packets per second. However, the deployed model in the ISAC system may lack this capability, or the communication rate may be insufficient to support it. Consequently, the system is unable to process the user's request.

One effective way to tackle the challenges mentioned above is to use mixture of experts (MoE) architecture, which includes a gating network and several experts, e.g., sub-models, adept in various communication and sensing tasks. These experts can be deployed across ISAC devices. Upon receiving a user request, the gating function evaluates current communication conditions and input data to select the most suitable experts, thereby enhancing the comprehensiveness, flexibility, and scalability of ISAC service. The introduction of the MoE architecture in ISAC offers several advantages.

\begin{itemize}
\item \textbf{Specialization and Task Efficiency}: The MoE architecture distributes workload of an ISAC system across specialized experts, each tailored for specific sensing and communication tasks, hence enhancing the overall system performance and efficiency.

\item \textbf{Parallel Task Processing}: The MoE allows the ISAC system to employ multiple experts, each specializing in different areas. This enables it to handle multiple tasks, such as behavior and gesture recognition simultaneously, under various communication conditions, reducing system response latency.

\item \textbf{Enhanced Adaptability}: The MOE structure facilitates a flexible strategy for adding or removing experts. This allows the ISAC system to adjust the number and functionality of experts based on user communication and perception needs, thereby enhancing the system adaptability.

\end{itemize}

Given the challenges faced by ISAC and the potential benefits of MoE architecture, this article proposes an MoE enhanced ISAC framework, which includes multiple communication and sensing experts as well as a gating network. Upon receiving a user request, the gating network selects suitable experts based on current communication conditions and input data. Subsequently, the selected experts process the input independently to obtain the corresponding results. Finally, the gating network aggregates the results from multiple experts to produce the final output. Here, the ISAC experts could either focus on different communication and sensing tasks or be skilled in managing the same task across diverse situations, which can significantly enhance the overall system performance and robustness. The main contributions of this article are summarized as follows.
\begin{itemize}
    \item We analyze the potential support of MoE for ISAC systems from multiple perspectives, providing new insights for further optimizing ISAC via MoE to enhance system performance.

    \item  We propose the MoE enhanced ISAC framework that dynamically selects the suitable experts to handle users' requests by considering the communication conditions, inputs, and other relevant information, improving the system's performance and robustness.

    \item We evaluate the proposed framework through a case study of target detection based on Channel State Information (CSI). Experimental results demonstrate that, at low communication rates, the framework's detection accuracy exceeds that of individual experts by at least 18\%.
\end{itemize}

\section{Mixture of Experts and integrated sensing and communications}
This section first discusses ISAC. On this basis, we introduce MoE and explain how it enhances ISAC.

\subsection{Integrated sensing and communications }
ISAC is a paradigm that converges radar sensing and wireless communications into a single platform, aiming to simultaneously perform environmental sensing (target detection, speed estimation, environmental monitoring) and data communication~\cite{liu2022survey}. This integration enhances spectral efficiency and reduces hardware requirements, which is particularly crucial in the burgeoning fields of the Artificial Intelligence of Things (AIoT), Vehicle-to-everything (V2X), and smart homes. For instance, based on wireless communication signals, the authors in~\cite{wang2024acceleration} extracted characteristics of propagation path length changes and then used the support vector machine (SVM) to analyze these characteristics, thereby achieving fall detection in multi-target scenarios. Such methods reuse existing wireless communication signals for sensing, endowing wireless networks with expanded functions, which have broad application prospects. Based on the aforementioned principles and typical examples, we can see that ISAC holds the following key advantages.

\begin{itemize}

\item  \textbf{Resource Consumption}: Sharing infrastructure between sensing and communication modules reduces overall system costs and power demands, contributing to economic savings and sustainability.

\item  \textbf{Enhanced Functions}: ISAC uses shared resources to simultaneously execute sensing and communication functions, providing more comprehensive functionality and richer applications than systems designed solely for communication or sensing.

\item  \textbf{Coordination Gain}: Integrating sensing and communication allows two functions to reinforce each other. This symbiotic relationship allows both functions to benefit from each other, thereby enhancing the overall system performance.
\end{itemize}

Although ISAC shows great potential, it is still in the early stages of development and faces certain limitations as summarized below.

\begin{itemize}
    \item Due to limited computing and storage resources, integrating multiple models within a single ISAC device is challenging. This restricts ISAC's ability to meet diverse user needs for communication and sensing. Additionally, a limited number of models can compromise the overall accuracy and robustness of sensing and communication, as well as the ability to process tasks in a parallel manner.

    \item As application environments become increasingly complex, ISAC encounters scalability and adaptability challenges. Here, scalability means that ISAC systems need to continuously develop and improve to meet user demands, while adaptability requires the systems to dynamically adjust to changes in the external physical environment.
\end{itemize}

In response to these challenges, the MoE architecture offers an effective solution, which is detailed as follows.

\subsection{Mixture of Experts}
MoE is a machine learning framework, designed to manage complex tasks by breaking them into simpler, more manageable subtasks, each handled by a specialized expert. These experts, either models or algorithms, are trained to manage specific data aspects, thus providing detailed task insights.~\cite{yuksel2012twenty}. The core of MoE is the gating network, which dynamically directs input data to appropriate experts and combines their outputs to produce the final result. This modular and decentralized architecture offers the following advantages.
    \begin{figure*}[t]
	\centering
	\includegraphics[width=1\textwidth]{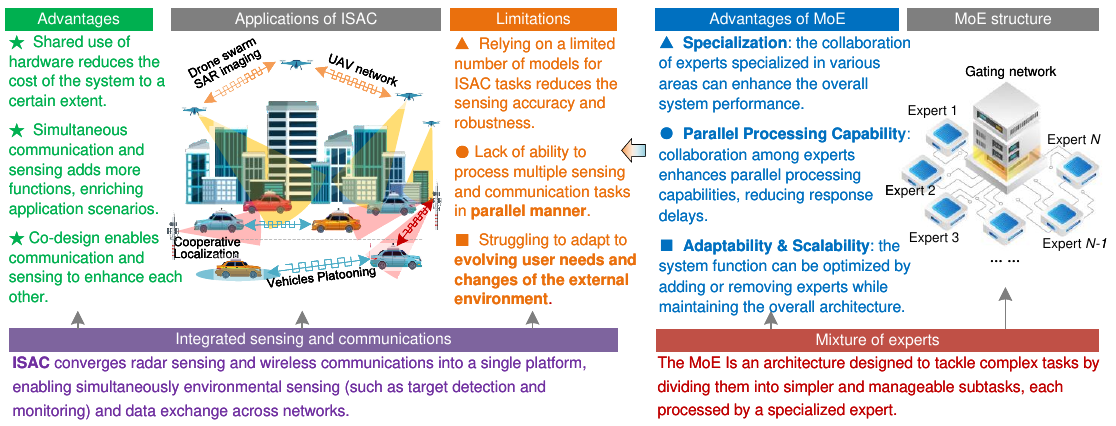}%
	\caption{An overview of ISAC, MoE, and their potential interconnections.}
	\label{ISACnMoE}
\end{figure*}
\begin{itemize}
    \item \textbf{Specialization}: MoE breaks down complex tasks into simpler subtasks, each handled by experts specialized in those specific areas, leading to improved overall performance. Integrating MoE into ISAC can facilitate the collaboration of multiple sensing and communication modules distributed across various devices, thereby enhancing the system performance. For example, in gesture recognition~\cite{zhang2018crosssense}, we can deploy different signal feature extraction algorithms and machine learning models across various ISAC devices. By selectively activating and combining these algorithms and models, we can achieve more accurate and robust gesture recognition.

     \item  \textbf{Parallel Processing}: The parallel processing capability of MoE significantly reduces the computing time, making them suitable for real-time applications. This method allows an ISAC system to manage multiple tasks at once, thereby providing support for applications such as digital twins and the Metaverse, which demand constant and real-time updates. For example, an ISAC system with localization and behavior recognition capabilities can activate and run multiple experts in parallel to simultaneously provide users with both services~\cite{gao2017csi}. However, given the limited computing and storage resources of a single ISAC device, models need to be distributed across different devices. Therefore, enhancing parallel processing requires increased information exchange among ISAC devices to ensure timely output aggregation.

    \item  \textbf{Adaptability}: The modular design of MoE supports scalability, enabling adjustments to the system’s function and capacity by adding or removing experts without major architectural changes. This allows ISAC to modify the types and numbers of models while preserving the overall structure, essential for the evolving of sensing and communication over time. For example, in a gesture recognition ISAC system based on the MoE, we can integrate experts in both localization and activity recognition without modifying the core system architecture. This enhances the ISAC system's capabilities, allowing it to meet users' dynamically changing needs. Note that adjustments to sensing models necessitate further optimization of the gating network to ensure objective expert selection.

\end{itemize}

In Fig.~\ref{ISACnMoE}, we summarize ISAC and MoE as well as the relationship between them.
\vspace{-0.5cm}

\section{Applications of MoE in ISAC}
This section discusses the MoE’s potential applications in ISAC from the signal processing, network optimization, and application perspectives.
\subsection{MoE for Signal Processing in ISAC}
In various application scenarios, ISAC requires different modulation methods to optimize both communication and sensing. The orthogonal frequency division multiplexing (OFDM) is the dominant modulation technology in existing communication systems, supporting both communication and sensing functions. However, in high-mobility environments, Doppler spread can cause severe inter-carrier interference (ICI), which impairs both sensing and communication performance. To overcome this issue, researchers proposed orthogonal time frequency space (OTFS) modulation. In OTFS, data symbols are mapped onto the delay-Doppler (DD) domain, where time-varying channels appear quasi-static and sparse. This mapping reflects the scatterers' geometry in high-mobility environments, making OTFS well-suited for wireless sensing~\cite{wei2022orthogonal}. Modulation, along with other signal processing modules, such as signal detection and synchronization, are shared components between sensing and communication. They have varying functional preferences, where some are better for communication, while others are more suited for sensing. Therefore, integrating modules (such as OFDM and OTFS) through the MoE architecture allows ISAC to selectively activate these modules as needed. This strategic activation maximizes the strengths of each module, thereby enhancing overall ISAC performance.

In addition to the shared modules mentioned above, considering sensing and communication as two functions of ISAC the MoE is also indispensable. For wireless communication, MoE-AMC~\cite{gao2023moe} combines a multilayer perceptron based gating network with specialized experts to achieve automatic modulation classification (AMC). It features a recognition model based on ResNet for signals with high Signal-to-Noise Ratio (SNR) and a Transformer-based model for signals with low SNR. During operation, the gating network evaluates the received signal's SNR, activates the appropriate expert, and combines their outputs to effectively perform AMC. For sensing tasks such as behavior and gesture recognition, MoE is even more crucial due to the diversity of functions and signal processing demands. For example, ISAC-based fall detection uses FFT for time-frequency analysis, while localization tasks involve extracting spatial features like flight time for geometric constraints~\cite{wang2024through}. Therefore, from the sensing perspective, MoE is also indispensable as it can integrate various signal feature extraction algorithms, such wavelet analysis and fractional Fourier transform, and selectively activate experts based on the requirements, communication rate, and computing resources, achieving more efficient sensing.

\subsection{MoE for Resource Optimization in ISAC}
In ISAC, resource optimization in ISAC uses standard performance metrics for communication, but sensing involves varied functions with diverse evaluation criteria. For instance, localization relies on array aperture and signal bandwidth, and its performance is measured by distance error, while behavior recognition depends on communication rates and is assessed by recognition accuracy, such as true positive rate. This diversity poses challenges for resource optimization within the ISAC framework. In response, in~\cite{dong2022sensing}, the authors considered various sensing quality of service metrics, including the Cramér-Rao Lower Bound (CRLB) for signal parameter estimation and the probability of target detection. They established a framework to allocate power and bandwidth resources for both sensing and communication using convex optimization. In another study~\cite{wang2023resource}, a down-link multi-cell ISAC system was designed to concurrently manage communication and multiple target location estimation. This system uses a deep reinforcement learning (DRL)-based strategy for sub-channel and power allocation, aiming to maximize total communication rate and ensure minimum SINR for each user, while optimizing the CRLB for accurate localization. As can be observed, ISAC system requires different optimization strategies for various cases. Therefore, using MoE to integrate resource allocation solutions tailored for different ISAC scenarios is necessary. In such context, sensing and communication requirements, along with available resources such as bandwidth and power, serve as inputs to the gating network, which then selects the suitable expert for achieving the optimization solution.

For optimal resource allocation, Generative AI (GAI) serves as an effective approach in addition to traditional methods such as convex optimization and DRL. For instance, in~\cite{wang2024unified}, the authors proposed WiPe-GAI, a framework that uses wireless sensing to guide digital content generation. To ensure the effectiveness of the proposed framework in resource-constrained networks, a pricing-based incentive mechanism is designed, and a diffusion model-based approach is introduced to generate optimal pricing strategy. This strategy aims to maximize user utility while incentivizing virtual service providers to participate in service provisioning. However, GAI models are typically large in scale, and their training and inference processes are resource-intensive. Meanwhile, ISAC features diverse performance metrics and application scenarios. Hence, integrating MoE is essential when utilizing GAI for ISAC resource optimization, which mainly includes the following two aspects.

\begin{enumerate}

    \item From the perspective of GAI models, MoE can enhance their structure~\footnote{https://community.openai.com/t/gpt-4-cost-estimate-updated/578008}, reducing the energy consumption and complexity during the training and inference process, thereby increasing optimization efficiency.

    \item MoE can integrate multiple GAI-based resource optimization experts to solve a variety of optimization problems in different scenarios.

\end{enumerate}

Thereby, given the comprehensive functions, diverse evaluation metrics, and broad application scenarios, integrating MoE into ISAC resource allocation process is essential for improving the optimization efficiency.

\subsection{MoE for ISAC Applications}
Besides GAI, DAI models, such as SVM and Long short-term memory (LSTM) are also crucial, especially for the application of ISAC. Unlike GAI, which excels in data generation, DAI models are commonly used for data analysis, classification, and prediction. In ISAC, DAI can not only support communication-related tasks such as channel estimation but also analyze various signal characteristics, thereby facilitating different sensing tasks. For instance, the authors in~\cite{nguyen2021svm} treated a channel estimation problem for uncorrelated channels as a conventional SVM problem and modified the objective function to estimate spatially correlated channels. On this basis, they proposed an SVM based joint channel estimation and data detection method, which leverages both to-be-decoded data and pilot data to enhance estimation and detection performance. For sensing, the authors in~\cite{chen2018wifi} introduced an attention-based bi-directional long short-term memory (ABLSTM) approach. It learns features in both directions from the original sequential CSI measurements. Then, they further employed an attention mechanism to assign varying weights to all the learned features, thereby achieving better human activity recognition performance.

\begin{figure*}[t]
	\centering
	\includegraphics[width=1\textwidth]{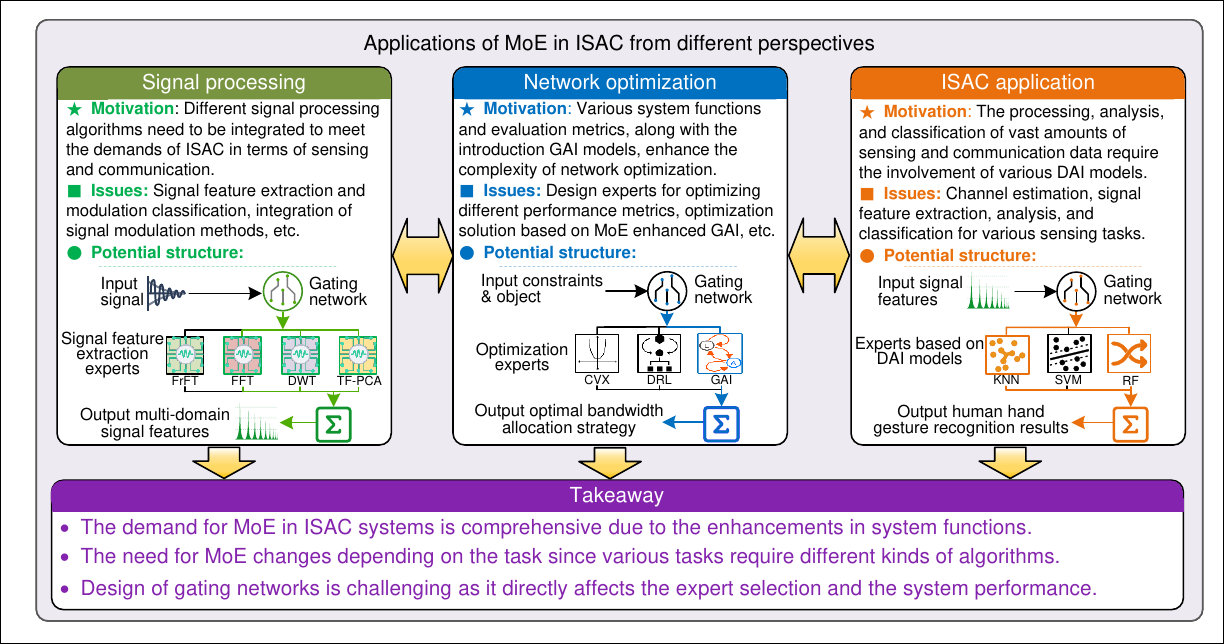}%
	\caption{Summary of MoE applications in ISAC systems. In signal processing, the FrFt, FFT, DWT, and TF-PCA indicate Fractional Fourier transform, fast Fourier transform, discrete wavelet transform, and time-frequency principal component analysis, respectively. In network optimization, CVX and DRL are convex and deep reinforcement learning, respectively. In applications, KNN, SVM, and RF mean K-Nearest neighbours, supported vector machine, and random forest, respectively. }
 \vspace{-0.5cm}
	\label{APLCT}
\end{figure*}

As discussed, due to the sensing and communication capabilities, ISAC has a strong demand for DAI models. Particularly for sensing, different purposes require various DAI models to analyze and classify signal features. For instance, binary classifiers including SVM are suitable for tasks such as detecting human targets, whereas multi-class classifiers such as random forests and Bayesian classifiers are better for gesture and behavior recognition. In this context, a crucial aspect is the compatibility between signal features and classifiers, which is key to the sensing performance. An example is the CrossSense~\cite{zhang2018crosssense}, a gesture recognition system based on WiFi. CrossSense uses multiple gesture recognition experts, each optimized for specific signal features and classifiers. It employs a K-Nearest Neighbors (KNN) based gating network to select the experts best suited to the input signal for gesture recognition. This improves the system's recognition accuracy and its robustness to environmental variations. Therefore, using MoE to integrate various DAI models is a promising approach to advancing ISAC system applications. This integration structure can be multi-tiered. For DAI models commonly used in both communication and sensing, we can designate them as shared experts, while other DAI models can be individually designed for the two modules. On this basis, it is also necessary to set up different gating networks for sensing and communication to activate experts based on specific requirements. Figure~\ref{APLCT} summarizes the application of MoE in ISAC.

\subsection{Lessons Learned}
Based on the discussion of the above three aspects, the following key insights can be summarized.

\begin{itemize}
\item \textbf{Comprehensive Demand for MoE in ISAC Systems.} This demand is primarily driven by enhancements in system functions. Concretely, ISAC systems offer more functions compared to traditional systems focused solely on communication or sensing. While sensing and communication share some system resources, their principles are fundamentally different, leading to more diverse and complex computing requirements. Additionally, these enhanced functions open up a broader range of applications,  requiring ISAC to adapt continually to external conditions, further increasing the system complexity. This underscores the pressing need for MoE.

\item \textbf{Varied Demand for MoE Across Different Aspects.}  This variation stems from the different model requirements of various ISAC tasks. For instance, ISAC physical layer tasks need signal processing algorithms, which are relatively fixed, predictable and controllable in terms of computation, energy consumption, and timing. In contrast, network optimization tasks often involve more complex AI algorithms that require greater computing resources and vary according to different constraints and objectives. Hence, the network optimization and application tend to have a greater need for MoE.

\item \textbf{Challenges in Designing Gating Networks.} Implementing MoE in ISAC systems allows for dynamic adjustments in the function and number of experts based on needs, enhancing system robustness and scalability. However, designing effective gating networks is a significant challenge, as these networks are crucial for reliable expert selection and hence directly impact overall system performance. With a limited pool of experts, a well-performing gating network can resolve more problems effectively by  leveraging the strengths of various experts.

\end{itemize}

\section{MoE Enhanced ISAC }
This section introduces an ISAC framework based on MoE for detecting the number of targets under various communication conditions. After that, we evaluate the proposed framework through a case study.
 \subsection{Motivation}
Detecting the number of dynamic human targets in specific areas is a foundational function of CSI based ISAC systems. This function supports the monitoring and control of human flow, presenting a substantial application market in settings such as airports and shopping centers. Existing CSI-based ISAC systems effectively detect the number of targets using machine learning to analyze CSI characteristics across multiple domains. However, these systems exhibit two limitations, which hinder their application in practical scenarios.

 \begin{itemize}
\item Existing works focus on the detection performance while neglecting the demands on communication rates. In practice, when users initiate requests, the communication rate of ISAC systems may be low, making it challenging to execute detection. In such cases, the system performance could markedly decrease or even fail to operate properly.

\item Due to constraints such as limited signal bandwidth, the signal features, used by existing systems, have limited resolution. Meanwhile, various machine learning algorithms exhibit different sensitivities to signal features. These factors cause significant fluctuations in detection performance as the external conditions, such as the number of targets, shift.
\end{itemize}

One solution is to use MoE to integrate multiple detection experts across various ISAC devices, enhancing detection capabilities. While all experts aim to detect the number of targets, they utilize different signal features and machine learning methodologies and have varying requirements for communication rates. Consequently, the gating network can activate suitable experts based on the input and current communication state to enhance the system's detection performance and robustness.
\begin{figure*}[t]
	\centering
\includegraphics[width=1\textwidth]{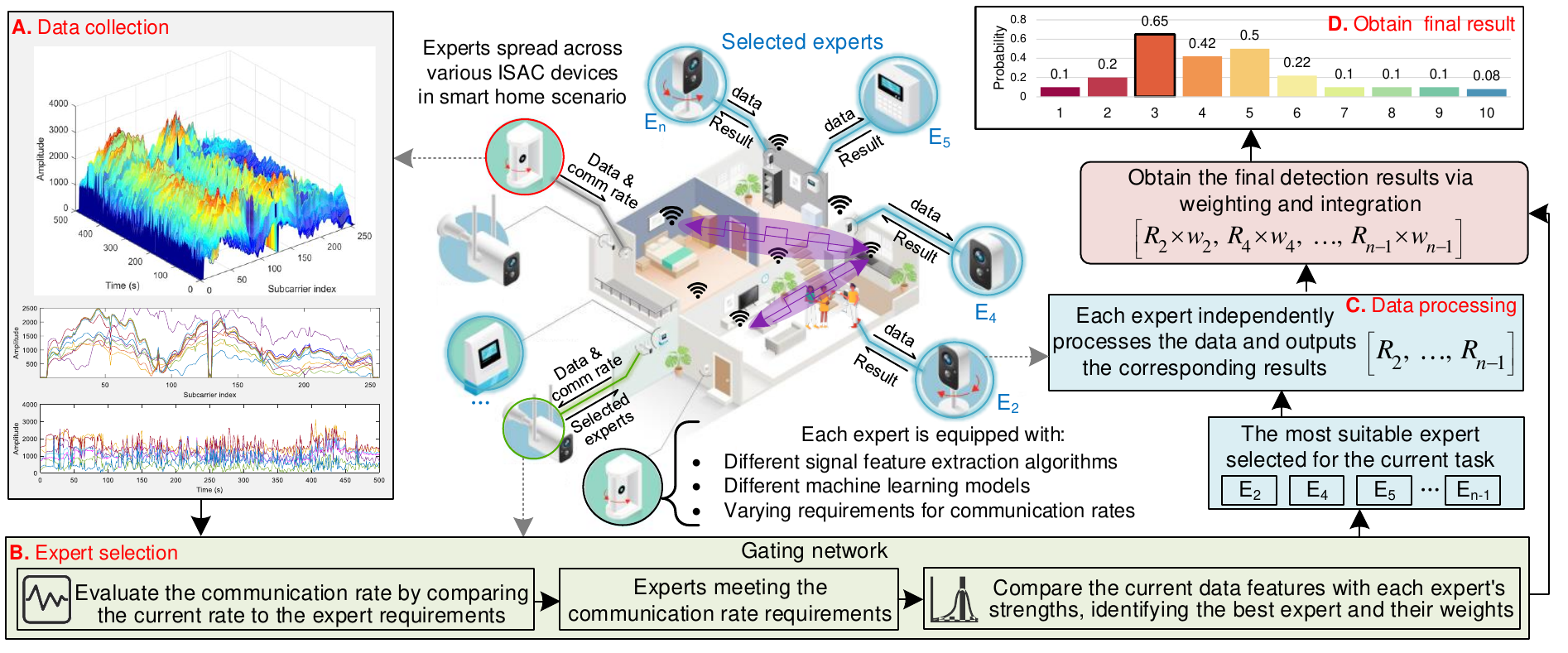}%
	\caption{The structure of the proposed framework. While this paper uses detecting the number of targets as an example, the framework can incorporate experts with different functions. This integration allows for a broader range of sensing tasks to be completed under various communication conditions, providing users with more comprehensive services.}
 \vspace{-0.5cm}
	\label{FW}
\end{figure*}
\subsection{The Proposed Framework}
We propose an MoE-enhanced framework consisting of multiple detection experts across ISAC devices and a gating network to detect human targets in monitored areas. Each expert utilizes CSI feature extraction combined with machine learning for detection. They differ in specific features and machine learning models and have varying communication rate requirements. The gating network selects experts based on the CSI input and current communication rate, processes the input, and aggregates their results for detection. This framework ensures experts operate effectively under current conditions and enhances system robustness by integrating multiple experts' outputs. The general working processes of this framework are outlined as follows.

\begin{enumerate}
    \item When a user requests detection, the gating network evaluates the current communication rate and expert requirements across ISAC devices, identifying those that can operate under current conditions.

    \item As different experts specialize in distinct signal features, on the basis of the first step, the gating network then evaluates them based on the input CSI characteristics to determine the most suitable experts for processing the current CSI input.

    \item The selected experts process the collected CSI data to obtain the detection results. Since they use different CSI features and machine learning methods, the detection outputs from each expert may be different.

    \item The detection outputs from experts deployed on different ISAC devices are weighted and integrated, yielding the final detection result which is then provided to the user.

\end{enumerate}

Figure~\ref{FW} illustrates the workflow of the proposed framework. It is worth noting that the proposed framework is explained through the example of detecting the number of targets. However, as shown in the Fig.~\ref{FW}, it is fundamentally capable of integrating models with various functions, including behavior recognition, gesture recognition, and localization. Therefore, when the local ISAC device cannot handle the user's sensing request, it can select models deployed on other devices to fulfil the task, offering users a broad range of ISAC services.

\subsection{Case Study}
\subsubsection{Experimental Configurations} 
\begin{figure*}[t]
	\centering
\includegraphics[width=1\textwidth]{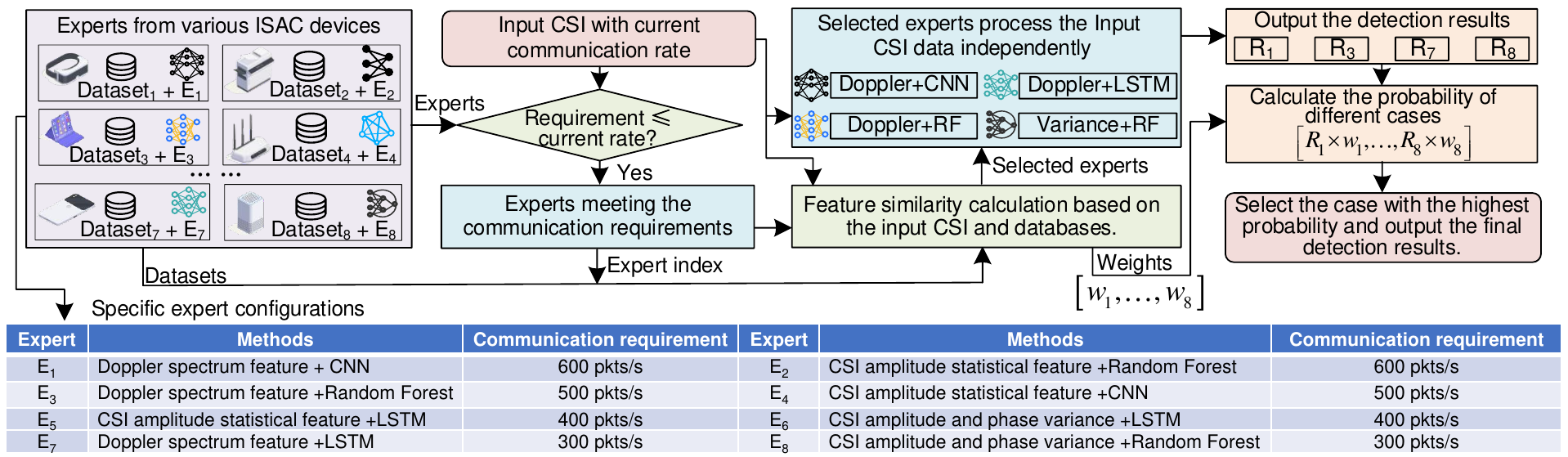}%
	\caption{The detection process based on the proposed framework and the expert configuration. Here, the Doppler feature refers to the energy distribution across various frequencies. The statistical features of CSI amplitude include variance, absolute average deviation, median, quartiles, and so forth.}
	\label{case}
\end{figure*}

Consider a MoE based ISAC framework for detecting the number of targets under varying communication scenarios. The framework includes a gating network and eight target detection experts. These experts employ different signal features and machine learning models for target detection and hold different communication rate requirements, as illustrated in Fig.~\ref{case}. Additionally, each expert is linked to a template dataset where it shows good detection performance.

During operation, the gating network first selects experts that can operate effectively at the current communication rate. For example, if the current rate is 600 packets per second (pkts/s), then only experts requiring a transmission rate of 600 or less can be chosen. Subsequently, the gating network extracts signal features from the input CSI and perform correlation calculations between the extracted features and those of the template dataset corresponding to the selected experts, identifying three most suitable experts. After that, these selected experts process the input CSI data to obtain the detection results. Finally, the gating network employs the computed correlations to perform a weighted fusion of the output results from the experts, yielding the final detection result. Noted that if the current communication rate is too low for any experts to participate, then the gating network directly selects experts through correlation calculations, ensuring the system operates effectively.

\subsubsection{Performance Analysis} 
\begin{figure}[tbp!]
  \centering
  \includegraphics[height=7.5 cm]{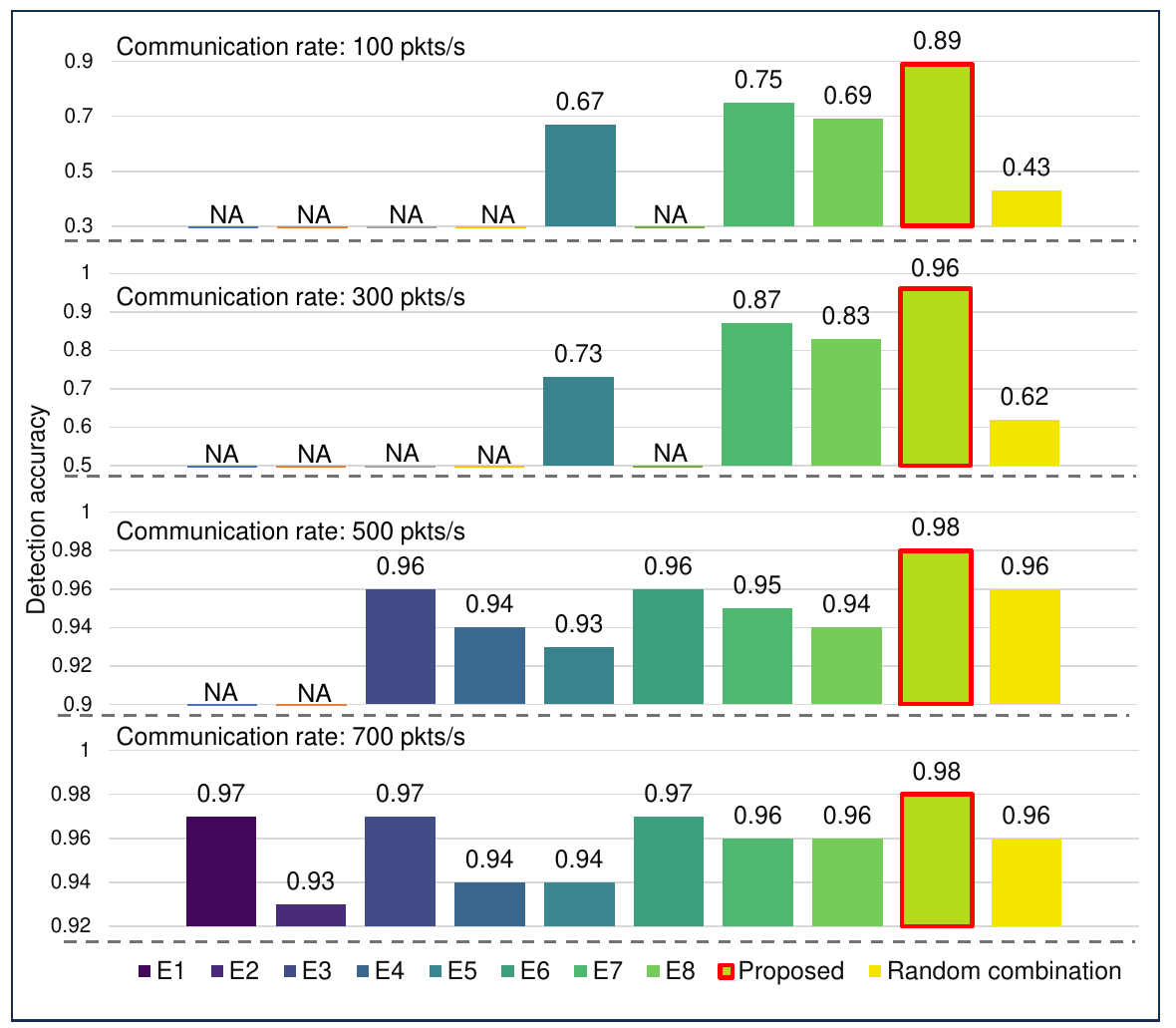}
  \caption{The detection results under cases of different communication rates. In cases with 100 pkts/s and 300 pkts/s, the proposed method activates $E_5$, $E_7$, and $E_8$ for detection via correlation calculations. Consequently, as a comparison, we present the detection accuracy of each of these experts individually. }
  \vspace{-1cm}
  \label{RST1}
\end{figure}

First, we evaluate the proposed framework at various communication rates. Figure~\ref{RST1} displays the detection accuracies for experts meeting communication requirements, the proposed framework, and three randomly combined experts. As can be seen, at various communication rates, the proposed framework outperforms others. For example, with a transmission rate of 500 pkts/s, the accuracies of $E_3$, $E_4$, $E_5$, $E_6$, $E_7$, and $E_8$ are 0.96, 0.94, 0.93, 0.96, 0.95, and 0.94, respectively, while combining three experts randomly results in a detection accuracy of 0.96. In contrast, the proposed framework achieves 0.98, outperforming the other approaches. When the transmission rate drops to 100 pkts/s, the proposed framework activates $E_5$, $E_7$, and $E_8$ for detection, achieving an detection accuracy of up to 0.89, which is higher than that obtained by each expert individually or by randomly combining experts. These results confirm that the effectiveness of the designed gating network, i.e., it can select the most suitable experts based on the current communication status and input, and then fuse their outputs for detection. Additionally, the use of the MoE enhances the system's overall detection accuracy across different communication rates, offering insights for further improving the performance of ISAC systems.

\begin{figure}[tbp!]
  \centering
  \includegraphics[height=5cm]{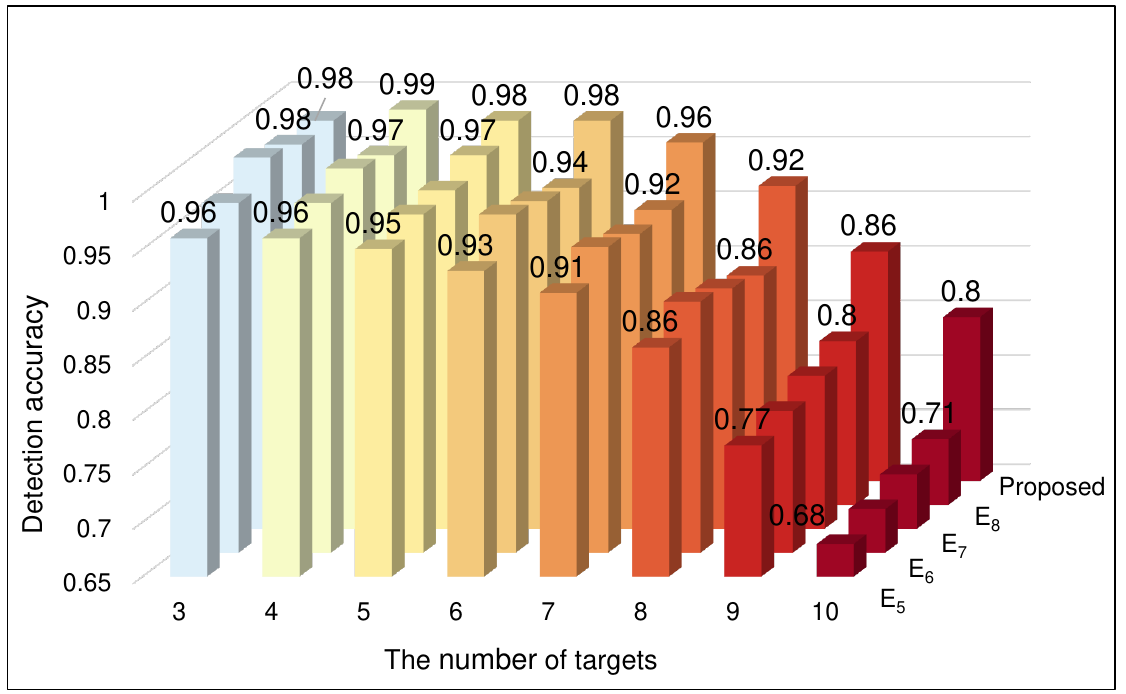}
  \caption{The detection results in cases with different number of targets.}
 \vspace{-0.5cm}
  \label{RST2}
\end{figure}

Then, we conduct tests with different number of targets to assess the framework's robustness. The results in Fig.~\ref{RST2} indicate that the detection performance of each expert declines as the number of targets increases. For instance, when the number of targets rises from 3 to 10, the accuracy of $E_5$ drops from 0.96 to 0.68 and that of $E_8$ decreases from 0.98 to 0.71. Meanwhile, our framework only declines from 0.98 to 0.80, which is better than that of $E_5$ and $E_8$. This is attributed to our framework activates multiple experts for detection and fuses their outputs to form the final result, enhancing its overall robustness. Therefore, we can see that integrating the MoE into ISAC can not only improve the system performance but also the robustness, which further improves the ISAC system and facilitates its practical deployment and application.
\section{Future Directions}
\subsection{ISAC Security}
The MoE based ISAC can be optimized by adjusting the number and function of experts. To ensure the effective operation after adjustments, the gating network needs to be upgraded. Therefore, designing an efficient mechanism to update gating networks is crucial. During design, it is essential to consider the expert functions, communication requirements, and the relationships between the experts and current gating strategies. Additionally, the constraints of newly added experts must also be considered. This will ensure that the updated gating network can balance both existing and new experts, thereby selecting the most suitable experts for new tasks and facilitating the functional evolution of the ISAC system.
\subsection{Resource Optimization}
MoE enhances the performance of ISAC systems by integrating multiple experts, but this results in increased computing and transmission resource consumption. Different experts may operate in varied environments, leading to different costs. Therefore, future work needs to further optimize MoE from a resource allocation perspective. For example, selecting experts requires considering their suitability for tasks and the computing and transmission resources they consume. This will enable the MoE based ISAC network to operate in a more economical manner.
\subsection{Interactive AI}
Besides ISAC, MoE is crucial for interactive AI systems. In the interaction between humans and AI models, the AI models need to understand, process, and predict the constantly changing needs of users. Therefore, it is vital to incorporate MoE into the construction of interactive AI models. This will enable the interactive AI model to invoke different experts based on user needs. Concurrently, new experts can be added to the existing model, which allows for the integration of new skills while retaining existing function, therefore achieving lifelong learning to meet the ever-changing demands of users.

\section{Conclusions}
This article proposes utilizing the MoE architecture to enhance the ISAC system, improving its sensing accuracy and robustness across various communication rates. Specifically, we first discuss ISAC and MoE, including concepts, advantages, and limitations. Building on this, we propose an MoE-enhanced ISAC framework that includes multiple experts and a gating network. The multiple experts are capable of performing various sensing tasks under different communication rates, while the gating network is designed to select the appropriate experts by analyzing current communication rates and input data characteristics. In this way, our framework can coordinate multiple experts to perform sensing under various communication conditions, thereby ensuring the performance. Our case study confirmed the framework's advancement under various communication conditions, offering insights for further optimizing ISAC systems.
\bibliographystyle{IEEEtran}
\bibliography{Ref.bib}

\begin{thebibliography}{10}
\providecommand{\url}[1]{#1}
\csname url@samestyle\endcsname
\providecommand{\newblock}{\relax}
\providecommand{\bibinfo}[2]{#2}
\providecommand{\BIBentrySTDinterwordspacing}{\spaceskip=0pt\relax}
\providecommand{\BIBentryALTinterwordstretchfactor}{4}
\providecommand{\BIBentryALTinterwordspacing}{\spaceskip=\fontdimen2\font plus
\BIBentryALTinterwordstretchfactor\fontdimen3\font minus
  \fontdimen4\font\relax}
\providecommand{\BIBforeignlanguage}[2]{{%
\expandafter\ifx\csname l@#1\endcsname\relax
\typeout{** WARNING: IEEEtran.bst: No hyphenation pattern has been}%
\typeout{** loaded for the language `#1'. Using the pattern for}%
\typeout{** the default language instead.}%
\else
\language=\csname l@#1\endcsname
\fi
#2}}
\providecommand{\BIBdecl}{\relax}
\BIBdecl

\bibitem{liu2022integrated}
F.~Liu, Y.~Cui, C.~Masouros, J.~Xu, T.~X. Han, Y.~C. Eldar, and S.~Buzzi,
  ``Integrated sensing and communications: Toward dual-functional wireless
  networks for 6g and beyond,'' \emph{IEEE journal on selected areas in
  communications}, vol.~40, no.~6, pp. 1728--1767, 2022.

\bibitem{wang2024generative}
J.~Wang, H.~Du, D.~Niyato, Z.~Xiong, J.~Kang, B.~Ai, Z.~Han, and D.~I. Kim,
  ``Generative artificial intelligence assisted wireless sensing: Human flow
  detection in practical communication environments,'' \emph{arXiv preprint
  arXiv:2404.14140}, 2024.

\bibitem{liu2022survey}
A.~Liu, Z.~Huang, M.~Li, Y.~Wan, W.~Li, T.~X. Han, C.~Liu, R.~Du, D.~K.~P. Tan,
  J.~Lu \emph{et~al.}, ``A survey on fundamental limits of integrated sensing
  and communication,'' \emph{IEEE Communications Surveys \& Tutorials},
  vol.~24, no.~2, pp. 994--1034, 2022.

\bibitem{wang2024acceleration}
J.~Wang, H.~Du, D.~Niyato, M.~Zhou, J.~Kang, and H.~V. Poor, ``Acceleration
  estimation of signal propagation path length changes for wireless sensing,''
  \emph{IEEE Transactions on Wireless Communications}, 2024.

\bibitem{yuksel2012twenty}
S.~E. Yuksel, J.~N. Wilson, and P.~D. Gader, ``Twenty years of mixture of
  experts,'' \emph{IEEE transactions on neural networks and learning systems},
  vol.~23, no.~8, pp. 1177--1193, 2012.

\bibitem{zhang2018crosssense}
J.~Zhang, Z.~Tang, M.~Li, D.~Fang, P.~Nurmi, and Z.~Wang, ``Crosssense: Towards
  cross-site and large-scale wifi sensing,'' in \emph{Proceedings of the 24th
  annual international conference on mobile computing and networking}, 2018,
  pp. 305--320.

\bibitem{gao2017csi}
Q.~Gao, J.~Wang, X.~Ma, X.~Feng, and H.~Wang, ``Csi-based device-free wireless
  localization and activity recognition using radio image features,''
  \emph{IEEE Transactions on Vehicular Technology}, vol.~66, no.~11, pp.
  10\,346--10\,356, 2017.

\bibitem{wei2022orthogonal}
Z.~Wei, S.~Li, W.~Yuan, R.~Schober, and G.~Caire, ``Orthogonal time frequency
  space modulation—part i: Fundamentals and challenges ahead,'' \emph{IEEE
  Communications Letters}, vol.~27, no.~1, pp. 4--8, 2022.

\bibitem{gao2023moe}
J.~Gao, Q.~Cao, and Y.~Chen, ``Moe-amc: Enhancing automatic modulation
  classification performance using mixture-of-experts,'' \emph{arXiv preprint
  arXiv:2312.02298}, 2023.

\bibitem{wang2024through}
J.~Wang, H.~Du, D.~Niyato, M.~Zhou, J.~Kang, Z.~Xiong, and A.~Jamalipour,
  ``Through the wall detection and localization of autonomous mobile device in
  indoor scenario,'' \emph{IEEE Journal on Selected Areas in Communications},
  vol.~42, no.~1, pp. 161--176, 2024.

\bibitem{dong2022sensing}
F.~Dong, F.~Liu, Y.~Cui, W.~Wang, K.~Han, and Z.~Wang, ``Sensing as a service
  in 6g perceptive networks: A unified framework for isac resource
  allocation,'' \emph{IEEE Transactions on Wireless Communications}, 2022.

\bibitem{wang2023resource}
X.~Wang, H.~Wu, Y.~Xu, H.~Cao, N.~Kumar, and J.~J. Rodrigues, ``Resource
  allocation in multi-cell integrated sensing and communication systems: A drl
  approach,'' in \emph{ICC 2023-IEEE International Conference on
  Communications}.\hskip 1em plus 0.5em minus 0.4em\relax IEEE, 2023, pp.
  3210--3215.

\bibitem{wang2024unified}
J.~Wang, H.~Du, D.~Niyato, J.~Kang, Z.~Xiong, D.~Rajan, S.~Mao, and X.~Shen,
  ``A unified framework for guiding generative ai with wireless perception in
  resource constrained mobile edge networks,'' \emph{IEEE Transactions on
  Mobile Computing}, 2024.

\bibitem{nguyen2021svm}
L.~V. Nguyen, A.~L. Swindlehurst, and D.~H. Nguyen, ``Svm-based channel
  estimation and data detection for one-bit massive mimo systems,'' \emph{IEEE
  Transactions on Signal Processing}, vol.~69, pp. 2086--2099, 2021.

\bibitem{chen2018wifi}
Z.~Chen, L.~Zhang, C.~Jiang, Z.~Cao, and W.~Cui, ``Wifi csi based passive human
  activity recognition using attention based blstm,'' \emph{IEEE Transactions
  on Mobile Computing}, vol.~18, no.~11, pp. 2714--2724, 2018.

\end{thebibliography}

\end{document}